\documentclass[a4paper,twocolumn,11pt,accepted=2017-05-17]{quantumarticle}
\pdfoutput=1
\usepackage[utf8]{inputenc}
\usepackage[english]{babel}
\usepackage[T1]{fontenc}
\usepackage{xcolor}
\usepackage{physics}
\usepackage{amsmath, amsthm, amssymb,mathrsfs,amsfonts}
\usepackage{multirow}
\usepackage[normalem]{ulem}
\newcommand{\stkout}[1]{\ifmmode\text{\sout{\ensuremath{#1}}}\else\sout{#1}\fi}
\usepackage{diagbox}
\usepackage{graphicx}
\usepackage{subfigure}
\usepackage{comment}
\usepackage{bbold}
\usepackage{hyperref}

\usepackage{tikz}
\usepackage{lipsum}

\newcommand{\jd}[1]{{\color{purple}#1}}

\usepackage[acronym]{glossaries}
\makeglossaries
\newacronym{QKD}{QKD}{quantum key distribution}
\newacronym{DI}{DI}{device-independent}
\newacronym{PM}{PM}{prepare-and-measure}
\newacronym{SDP}{SDP}{semi-definite programming}
\newacronym{QC}{QC}{quantum channel}

\begin{document}

\author{Marie Ioannou}
\affiliation{Department of Applied Physics University of Geneva, 1211 Geneva, Switzerland}
\author{Maria Ana Pereira}
\affiliation{Department of Applied Physics University of Geneva, 1211 Geneva, Switzerland}
\author{Davide Rusca}
\affiliation{Department of Applied Physics University of Geneva, 1211 Geneva, Switzerland}
\author{Fadri Grünenfelder}
\affiliation{Department of Applied Physics University of Geneva, 1211 Geneva, Switzerland}
\author{Alberto Boaron}
\affiliation{Department of Applied Physics University of Geneva, 1211 Geneva, Switzerland}
\author{Matthieu Perrenoud}
\affiliation{Department of Applied Physics University of Geneva, 1211 Geneva, Switzerland}
\author{Alastair A.\ Abbott}
\affiliation{Department of Applied Physics University of Geneva, 1211 Geneva, Switzerland}
\affiliation{Univ.\ Grenoble Alpes, Inria, 38000 Grenoble, France}
\author{Pavel Sekatski}
\affiliation{Department of Applied Physics University of Geneva, 1211 Geneva, Switzerland}
\author{Jean-Daniel Bancal}
\affiliation{Department of Applied Physics University of Geneva, 1211 Geneva, Switzerland}
\affiliation{Université Paris-Saclay, CEA, CNRS, Institut de physique théorique, 91191, Gif-sur-Yvette, France}
\author{Nicolas Maring}
\affiliation{Department of Applied Physics University of Geneva, 1211 Geneva, Switzerland}
\author{Hugo Zbinden}
\affiliation{Department of Applied Physics University of Geneva, 1211 Geneva, Switzerland}
\author{Nicolas Brunner}
\affiliation{Department of Applied Physics University of Geneva, 1211 Geneva, Switzerland}

\title{Receiver-Device-Independent Quantum Key Distribution} 

\begin{abstract}
We present protocols for quantum key distribution in a prepare-and-measure setup with an asymmetric level of trust. While the device of the sender (Alice) is partially characterized, the receiver's (Bob's) device is treated as a black-box. The security of the protocols is based on the assumption that Alice's prepared states have limited overlaps, but no explicit bound on the Hilbert space dimension is required. The protocols are immune to attacks on the receiver's device, such as blinding attacks. The users can establish a secret key while continuously monitoring the correct functioning of their devices through observed statistics. We report a proof-of-principle demonstration, involving mostly off-the-shelf equipment, as well as a high-efficiency superconducting nanowire detector. The possibility to establish a secret key is demonstrated over a $4.8\,$km low-loss optical fiber with a finite-size analysis. The prospects of implementing these protocols over longer distances is discussed.
\end{abstract}

\maketitle

\section{Introduction}

Quantum communication has witnessed an extremely fast evolution over the last two decades \cite{Diamanti2016,Xu2020,Pirandola2020}. On the practical level, record-distance implementations of \gls{QKD} have been reported \cite{Boaron2018, Yin2020,Chen2020}, and first commercial systems have been developed \cite{idq}. On the more fundamental level, significant progress has been achieved as well, notably through the development of the concept of \gls{DI} \gls{QKD} \cite{Acin2007,Pironio2009,Vazirani,Rotem2018}. The observation of strong nonlocal quantum correlations allows distant users to establish a secret in a black-box setting, i.e., without relying on a detailed characterization of their cryptographic devices. This represents the strongest form of security for \gls{QKD}~\cite{EkertRenner}. 

From a practical point of view, the concept of DI QKD has also generated interest, in particular as a potential solution for countering experimentally demonstrated hacking attacks~\cite{Lydersen2010,Gerhardt2011}. While first proof-of-principle experiments have just been reported using state-of-the-art setups \cite{nadlinger2021,zhang2021,liu2021}, any practical implementation of DI QKD is still arguably far out of reach.

This motivates research on more general scenarios for quantum communication where trust is relaxed on some of the observers or devices. The most well-known approach is that of Measurement-Device-Independent (MDI) QKD \cite{MDI,Braunstein12,Diamanti2016} where the honest parties (Alice and Bob) both send a quantum system to an intermediate third party (Charlie) performing a joint measurement. Security can be demonstrated without any assumption on Charlie's device, the protocol being in this sense MDI. However, the devices of both Alice and Bob must be characterized and cannot be treated as black boxes.

Another relevant scenario is the one where trust is relaxed on one of the honest parties. Consider for instance that Alice's device is trusted while Bob's device is viewed as a black-box. In practice, such an asymmetric scenario can be well-motivated, considering for example quantum communication between a large company and some end-user. On the one hand, the company has access to advanced technology and can verify the correct operation of its setup. On the other hand, the end user has only very limited resources and no possibility to verify the correct functioning of their cryptographic device. 

The above scenario, referred to as one-sided DI, has been introduced in \cite{Tomamichel2011}, and key rates have been derived considering the effect of noise and finite-size data \cite{Tomamichel,Tomamichel2017}. However, the effect of losses has not been considered in these works. Instead, a fair-sampling type assumption is made, which opens the door to the detection loophole and to attacks such as blinding \cite{Lydersen2010,Gerhardt2011}, which impose severe requirements in terms of transmission and detection efficiency \cite{Acin2016}. In practice, where losses are unavoidable, such an approach can thus no longer be considered one-sided DI. An alternative approach was followed in Ref.~\cite{Branciard}, considering an entanglement-based one-sided DI setup, establishing a connection to quantum steering. The implementation of such a setup is however challenging, as it requires a similar level of complexity compared to a fully DI protocol (notably in terms of detector efficiency) which explains why it has not been experimentally demonstrated so far.

In this work we discuss QKD protocols in a prepare-and-measure scenario where the sender's device is (partially) trusted while the receiver's device can be treated as a black-box. We term these protocols ``receiver-DI''. Our approach is based on the assumption that the prepared states have limited overlaps (i.e., we assume a bound on how distinguishable the states are from each other), inspired by recent developments in quantum randomness generation \cite{Brask2017,VanHimbeeck+17,Rusca+19} and quantum correlations in prepare-and-measure scenarios \cite{Wang2019}. Our approach can be classified as semi-DI \cite{Pawlowski2011} and one-sided DI \cite{Tomamichel2011}, but differs from previous proposals since (i) we do not need an explicit bound on the Hilbert space dimension of the quantum systems (as in Refs~\cite{Pawlowski2011,Woodhead2015}), and (ii) we do not rely on any type of fair-sampling assumption (as in Refs~\cite{Tomamichel,Tomamichel2017}).

In our protocols, the users can establish a secret key while continuously monitoring the quantum channel (as in any QKD protocol), but also continuously verifying (or self-testing) the correct operation of their devices. For Bob's device (and the communication channel) this verification procedure is performed based only on the observed statistics, similarly to the full DI model. Our protocols are therefore black-box on Bob's side, and do not rely on any physical model of the detector. This is a sensitive point in standard QKD protocols, because, contrary to the devices used for state preparation, the input of the detectors can be directly controlled by the eavesdropper Eve, as demonstrated in practice via so-called blinding attacks \cite{Lydersen2010,Gerhardt2011}. 
In contrast, our protocols are immune to attacks on Bob's device.

Alice's device requires a partial characterisation via the bounds on the overlap of the emitted states. We argue that this assumption can be rather well justified in practice, and our implementation features a monitoring module allowing Alice to ensure the validity of the bounds on the overlaps. In practice, the introduction of this additional assumption considerably simplifies the  implementation: a prepare-and-measure setup can be used, and low detection efficiencies can be tolerated (contrary to full DI  \cite{Pironio2009,Ma2011,Ho2020,Schwonnek2020,Woodhead2021,Sekatski2021} or entanglement-based one-sided DI protocols \cite{Branciard}).  

After presenting our protocols and their security analysis, we report on a proof-of-principle experiment. The setup involves mostly off-the-shelf equipment, with the addition of a high-efficiency superconducting nano-wire single-photon detector (SNSPD). For our simplest two-state protocol, an expected secret key rate of order $10^{-2}\sim 10^{-3}$ per round is demonstrated over a $4.8\,$km low-loss optical fiber, taking finite-size statistics into account. We illustrate the self-testing feature of the protocol, by showing that an artificial decrease of Bob's detector efficiency (as, e.g., in a blinding attack) is immediately detected by the users. Finally, we show that a three-state protocol can tolerate more loss than the two-state variation, and discuss more generally  the prospects of implementations over longer distances.

\section{Scenario}

\begin{figure}[b!]
    \centering
    \includegraphics[width=0.4\textwidth]{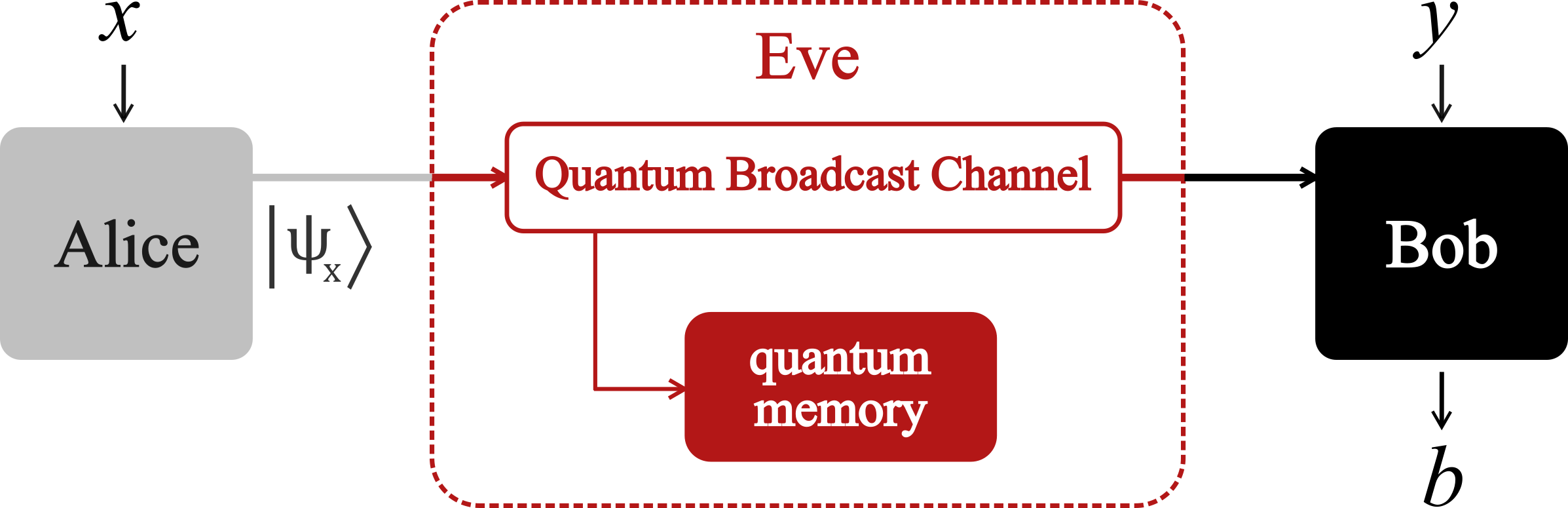}
    \caption{Scenario: Based on the observed data $p(b|x,y)$, and the assumption that Alice's preparations $\ket{\psi_x}$ have bounded overlap (see text), Alice and Bob can establish a secret key. Eve controls the quantum channel, but can also have full knowledge of the functioning of the devices of Alice and Bob.}
    \label{fig:scheme}
\end{figure}
	
We consider a prepare-and-measure setup, where Alice sends quantum systems to Bob who performs measurements (see Fig.~\ref{fig:scheme}). In each round, Alice prepares her system in one of the $n$ possible states $\{\ket{\psi_x}\}_{x=0}^{n-1}$ and sends it to Bob (note that the restriction to pure states is not necessary, see below). Bob then performs one of $n$ possible measurements, labelled by $y=0,...,n-1$, and obtains a binary outcome $b=0,1$. After sufficiently many repetitions, Alice and Bob can estimate the conditional probability distribution $p(b|x,y)$ \footnote{In practice to estimate the probability they reveal their outcomes on a small sample of the rounds chosen randomly, and use the results of the remaining rounds to distill the secret key.}. The security of the protocols, as given by a lower bound on the raw secret key rate (obtained after a post-processing step described below), is guaranteed based on the observed statistics $p(b|x,y)$, given that the setup complies with several assumptions that we now specify.

We begin with general assumptions common to all QKD protocols, including the DI case. (i) The choices of state preparation $x$ and measurement setting $y$ are made independently from Eve, i.e., she can not predict these values better than at random. (ii) No information about $x$ and $y$ leaks to Eve, except via the quantum and classical communication specified in the protocol at the given round. (iii) We assume the validity of quantum physics (note, that some DI protocols do not require this).  

 The central assumption specific to our protocols concerns the relation between the various states prepared by Alice. More precisely, (iv) we assume that their respective pairwise (possibly complex) overlaps $ \gamma_{ij} = \braket{\psi_i}{\psi_j} $ are bounded. One can think of the states $\ket{\psi_x}$ as describing the quantum system prepared by Alice's device and sent through the communication channel. More generally, they describe the states of all systems outside of Alice's lab conditioned on her applying the preparation sequence labeled by $x$. Note that if the states prepared by Alice are mixed, we require that they admit purifications that satisfy the overlap bounds -- the purifying system can be attributed to Alice's lab by assumption and does not compromise the security. The bound on the overlaps makes the no-leakage assumptions for $x$ in (ii) redundant; indeed it forbids the existence of a side-channel leaking any additional knowledge of $x$ to Eve. Nevertheless if such a side-channel is possible, it can be accounted for by adapting the overlaps $\gamma_{ij}$.

Note that we do not require to explicitly specify the relevant degrees of freedom or the Hilbert space dimension of the any quantum system. Loosely speaking, only the relative distinguishability of the states matters. Thus, Alice's device is partially characterized, but prone to unavoidable errors due to technical imperfections.

Concerning the receiver (Bob), no characterisation of their device is required and no fair-sampling type assumption is used. In particular, our protocols are robust to attacks where Eve controls Bob's device 
\cite{Lydersen2010}, which can compromise the security of standard \gls{QKD} protocols. This strong security comes however at a certain price, namely that the protocol is sensitive to losses. Importantly, this is not a particular weakness of our protocol, but a general feature of any QKD protocol that is device-independent on Bob's side (even considering a fully trusted Alice). Indeed, the possibility for Eve to perform a blinding attack sets severe bounds on the allowed transmission of the channel $\eta$. Specifically, no secret key can be obtained when $\eta \le 1/n$ \cite{Acin2016}; note that in order to overcome this limit, one may add the fair-sampling assumption, as in Ref.~\cite{Tomamichel}. As we will see below, our protocols can reach this limit (i.e.\ provide a positive key rate when $\eta \rightarrow 1/n$), and are therefore optimally robust to loss in the receiver-device-independent scenario.
 
Finally, note that in the present analysis we restrict the eavesdropper to collective attacks. That is, Eve interacts with each communication round independently of previous rounds and stores her systems in a quantum memory. We believe that our security analysis could be lifted to general attacks with established reduction techniques~\cite{renner2007,dupuis2016}.

\section{Protocols}

We now present our simplest protocol (where Alice can prepare $n=2$ different states), which is similar to the B92 protocol \cite{B92}. While the presentation below fits the implementation reported later, a more abstract and general presentation of these protocols is given in the companion paper \cite{Marie2}.  

Given a key bit $k=0,1$, Alice prepares one of two possible states, simply setting $x=k$. She uses a coherent state $\ket{\alpha} =e^{-|\alpha|^2/2} \sum_{n=0}^{\infty} \frac{\alpha^n}{\sqrt{n!}} \ket{n}$ with two possible polarization states $\ket{\phi_x}=\cos(\theta/2) \ket{H} + \sin(\theta/2)e^{i\pi x}\ket{V}$, which we write
\begin{equation}
     \ket{\psi_x}  = \ket{\alpha \cos(\theta/2)}_H  \ket{\alpha\sin(\theta/2)e^{i\pi x}}_V,
     \label{eq1}
\end{equation}
where $H $ and $V$ denote the two orthogonal polarization modes. The overlap between Alice's preparations is given by $\braket{\psi_1}{\psi_0} = e^{-2|\alpha|^2\sin(\theta)^2}$.
The main assumption of our protocol is then 
\begin{equation}
    \gamma = \braket{\psi_1}{\psi_0} \geq C,
    \label{eq3}
\end{equation}
where $C$ is a parameter chosen by the users.

Bob performs measurements of the polarization, using two possible bases. Specifically, for $y=0$, Bob projects the incoming signal onto a polarization $\ket{\phi_0^{\perp}}$, i.e., orthogonal to the polarization of Alice's first preparation. Similarly, for $y=1$, he projects onto a polarization $\ket{\phi_1^{\perp}}$. For both measurements, if Bob gets a click (i.e., detects some light in the orthogonal polarization mode), then the round is conclusive and he  outputs $b=0$; otherwise $b=1$, and the round will be discarded during sifting.

In the case of an ideal channel (loss and noise free), Alice and Bob will observe the statistics
\begin{equation}\label{eq4}
p(b=0|x,y) = 1-e^{-|\alpha|^2\sin(\theta)^2\sin(\frac{\pi(x-y)}{2})^2}.
\end{equation}
Note that $p(b=0|x,y)>0$ only when $x\neq y$. Hence, to establish the sifted key, Bob announces which rounds are successful, i.e.\ when $b=0$. In this case, Bob infers his raw key bit to be $k' = y \oplus 1$. For an ideal channel, Alice and Bob obtain a perfectly correlated sifted key, i.e.\ $k=k'$.

This protocol can be generalized to the case where Alice can prepare $n>2$ different states $\ket{\psi_x}$. In order to encode the raw key bit, Alice chooses now a pair of states, ${\bf r} =(r_0,r_1)$ with $0\leq r_0<r_1\leq n-1$, among ${n \choose 2}$ possible pairs. Then, for a key bit $k$, Alice sets $x=r_k$.
Note that every state $\ket{\psi_x}$ can now encode either bit value, $0$ or $1$. Bob has $n$ possible measurements, corresponding to  projections onto the polarizations orthogonal to the states that Alice can prepare. Each  measurement has two outputs $b=0,1$; $b=0$ corresponds to the projection onto $\ket{\phi_y^\perp}$ while $b=1$ corresponds to the projection onto $\ket{\phi_y}$. The sifting is now slightly more complicated. Alice announces ${\bf r}$, i.e., which pair of states she used. If Bob chose measurement $y=r_0$ or $y=r_1$ and got a conclusive outcome $b=0$, he announces that the round is successful; if not, the round is discarded.

For our implementation, we consider again polarized coherent states, similarly to Eq.~\eqref{eq1}, with $n$ possible polarizations $\ket{\phi_x} = \cos(\theta/2) \ket{H} + \sin(\theta/2)e^{ix2\pi/n}\ket{V}$, resulting in overlaps $ \gamma_{ij} = \braket{\psi_i}{\psi_j} = e^{-|\alpha|^2\sin(\theta/2)^2(1-e^{i2\pi/n(j-i)})}$. The main assumption in Eq.~\eqref{eq3} is now replaced by an assumption on these complex overlaps, i.e.\ the entries of the Gram matrix $G$ with elements $G_{ij}=\gamma_{ij}$. In fact, we can weaken this assumption by assuming only that the overlaps are in the vicinity of some ideal values $\gamma_{ij}$, i.e.\ $| \text{Re}(\gamma_{ij})- \text{Re}(\braket{\psi_i}{\psi_j}) | \leq \epsilon$ and $| \text{Im}(\gamma_{ij})- \text{Im}(\braket{\psi_i}{\psi_j}) | \leq \epsilon$ for a fixed $\epsilon$.

\section{Security analysis}

Alice and Bob must bound Eve's knowledge of the key in order to perform the final privacy amplification for distilling the secret key. In our protocol, this estimation can be performed based solely on the observed data $p(b|x,y)$, given the overlap assumption of Eq.~\eqref{eq3} is satisfied. 

The channel, controlled by Eve, is viewed as a quantum broadcast channel (see Fig.~1), where part of the information reaches Bob's lab, while the remainder is held by Eve. Given Alice sent the state $\ket*{\psi_{r_k}}$, the state at the output of the channel is denoted $\rho^{BE}_{r_k}$. Bob's measurements are denoted by a set of operators $M_{b|y}$. Eve performs a measurement on her subsystem, possibly after sifting, which is denoted $E_{e|z}$. As we do not impose any bound on the Hilbert space dimension, the measurements $M_{b|y}$ and $E_{e|z}$ can be taken to be projective (via Naimark dilation). 

Since the min-entropy lower-bounds the von Neuman entropy as $H_\text{min}(A|E,\text{succ}) \leq H(A|E,\text{succ})$, the asymptotic key rate (per round) is lower-bounded by~\cite{Devetak2005}
\begin{equation}
R = \left(H_\text{min}(A|E,\text{succ}) - H_2[\text{QBER}]\right)p({\text{succ}}),
\label{eq5}
\end{equation}
where $p(\text{succ})$ denotes the probability that a round of the protocol is conclusive (hence used to generate the key).
The second term captures the error correction cost ($H_2$ is the binary entropy and QBER is the quantum bit error rate, which can be estimated from the data $p(b|x,y)$). The first term captures the privacy amplification efficiency and is given by Eve's conditional min entropy $H_\text{min}(A|E,\text{succ})=-\log_2\left(p_g(e=k|\text{succ})\right)$, where $p_g(e=k|\text{succ})$ is the probability of correctly guessing the secret bit $k$ given that the round is conclusive. We note that the expression $R$ provides a lower bound on the key rate, and the bound is in general not tight. In order to obtain tight bounds, one should \sout{consider a different quantity, namely} directly consider the conditional von Neumann entropy, see e.g.\ Ref.~\cite{Devetak2005,dupuis2016,Fawzi2018}.

The main challenge is now to bound Eve's guessing probability $p_g :=p_g(e=k|\text{succ})= \frac{p(e=k,\text{succ})}{p(\text{succ})}$, under the constraint that the observed data is given by $p(b|x,y)$ and that the states prepared by Alice have given overlaps. More precisely, we want to upper bound the quantity: 
\begin{equation}
p_g = \frac{\sum_\mathbf{r} \sum_{k}  \trace{(\rho_{r_k}^{BE}(M_{0|r_0}+M_{0|r_1}) \otimes E_{k|\mathbf{r}})}}{\sum_\mathbf{r} \sum_{k} \trace{(\rho_{r_k}^{BE}(M_{0|r_0}+M_{0|r_1})\otimes \mathbb{1}_E)}}
\label{p_g}
\end{equation}
over any possible output state $\rho^{BE}_{r_k}$ and measurements for Bob and Eve that are compatible with the data and the overlap assumption. We assume that Alice and Bob choose their respective inputs ($k$, $\mathbf{r}$ and $y$) uniformly at random. It turns out that this problem can be relaxed to a hierarchy of semi-definite programs (SDPs) by taking advantage of the method introduced in Ref.~\cite{Wang2019} (see Appendix~\ref{EvesInfoApp}).

In Fig.~\ref{fig:theory_etaVSK}, we show how the corresponding bound on the key rate $R$ behaves as a function of the transmission $\eta$ of the channel, for protocols involving $n=2,3,4$ states. As noted before, any \gls{QKD} protocol in the receiver-DI model must have $R \rightarrow 0$ when $\eta \rightarrow 1/n$. 
Nevertheless, we see from the plot that the security approaches this threshold, so that lower transmissions can be reached by increasing the number of states $n$ in the protocol.

	\begin{figure}[t!]
    		\centering
    		\includegraphics[width=.5\textwidth]{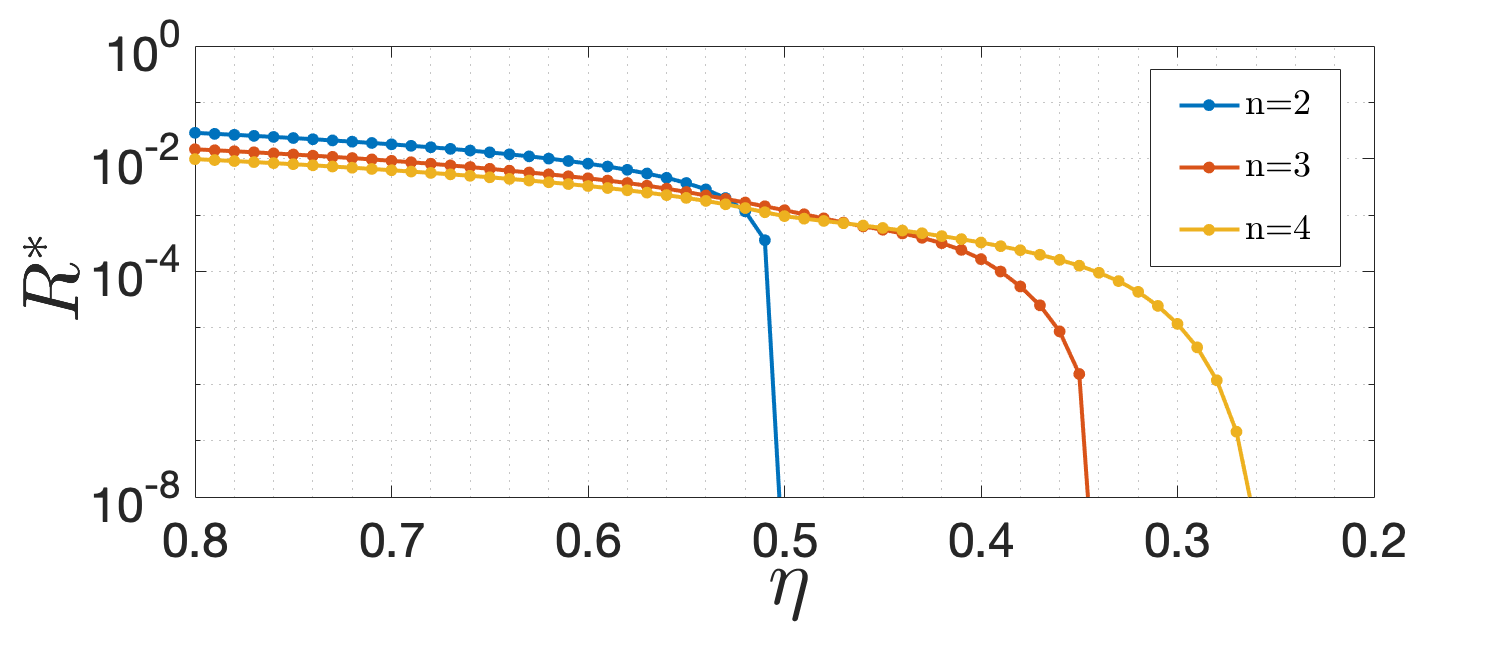}
    		\vspace{-10mm}
    		\caption{Lower bound, $R^*$, on the secret key rate $R$ as a function of the transmission $\eta$. A positive key rate is obtained for transmissions down to $\eta=1/n$, when considering a protocol where Alice prepares $n$ states of the form in Eq.~\eqref{eq1} (fixing the polarizations  to $\theta=0.2$ and optimizing over the coherent state $\alpha$).}
    		\label{fig:theory_etaVSK}
	\end{figure}

\section{Experimental realization}

	\begin{figure}[b!]
    		\centering
    		\includegraphics[width=.5\textwidth]{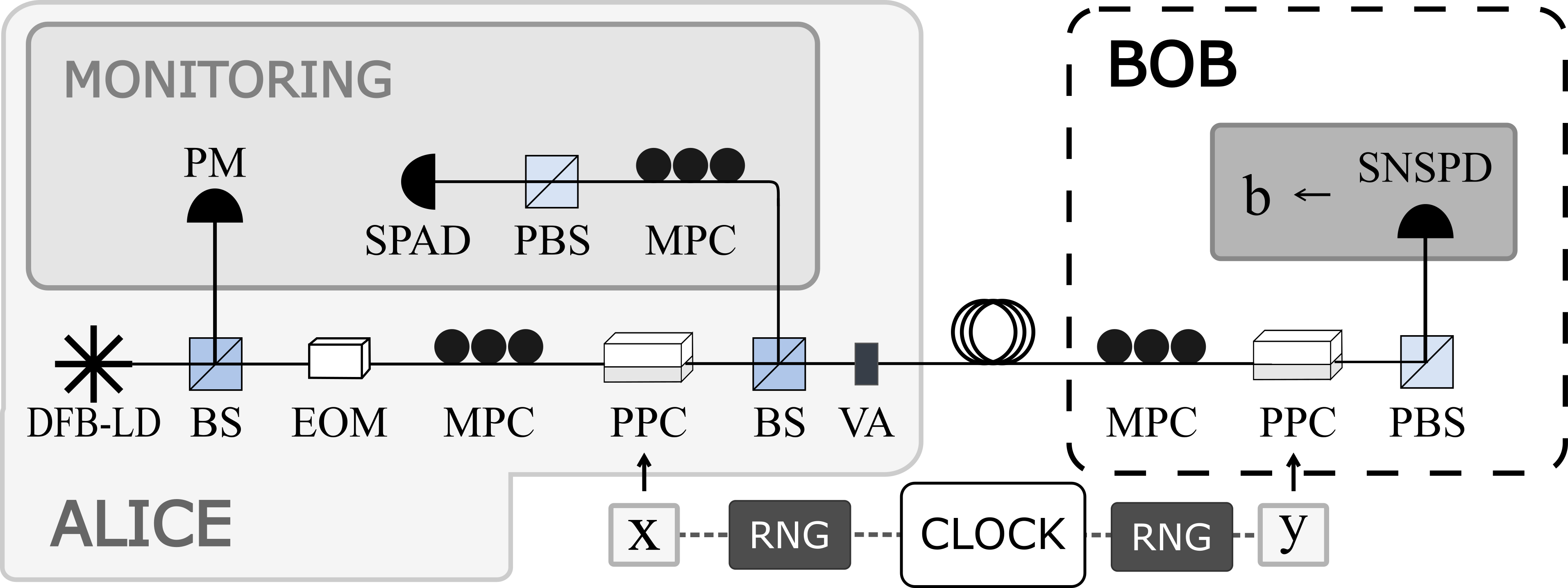}
    		\caption{Experimental setup. DFB-LD: Distributed feedback laser diode; BS: Beam splitter; EOM: Electro-optic modulator; MPC: Manual polarization controller; PPC: Piezo-electric polarization controller; VA: Variable attenuator; PBS: Polarizing beam splitter; SNSPD: Superconducting nanowire single-photon detector; SPAD: Single-photon avalanche diode; PM: Powermeter; RNG: Random number generator.}
    		\label{fig:exp_setup}
	\end{figure}

\begin{figure}[t]
    		\centering
	    \includegraphics[width=1.\linewidth]{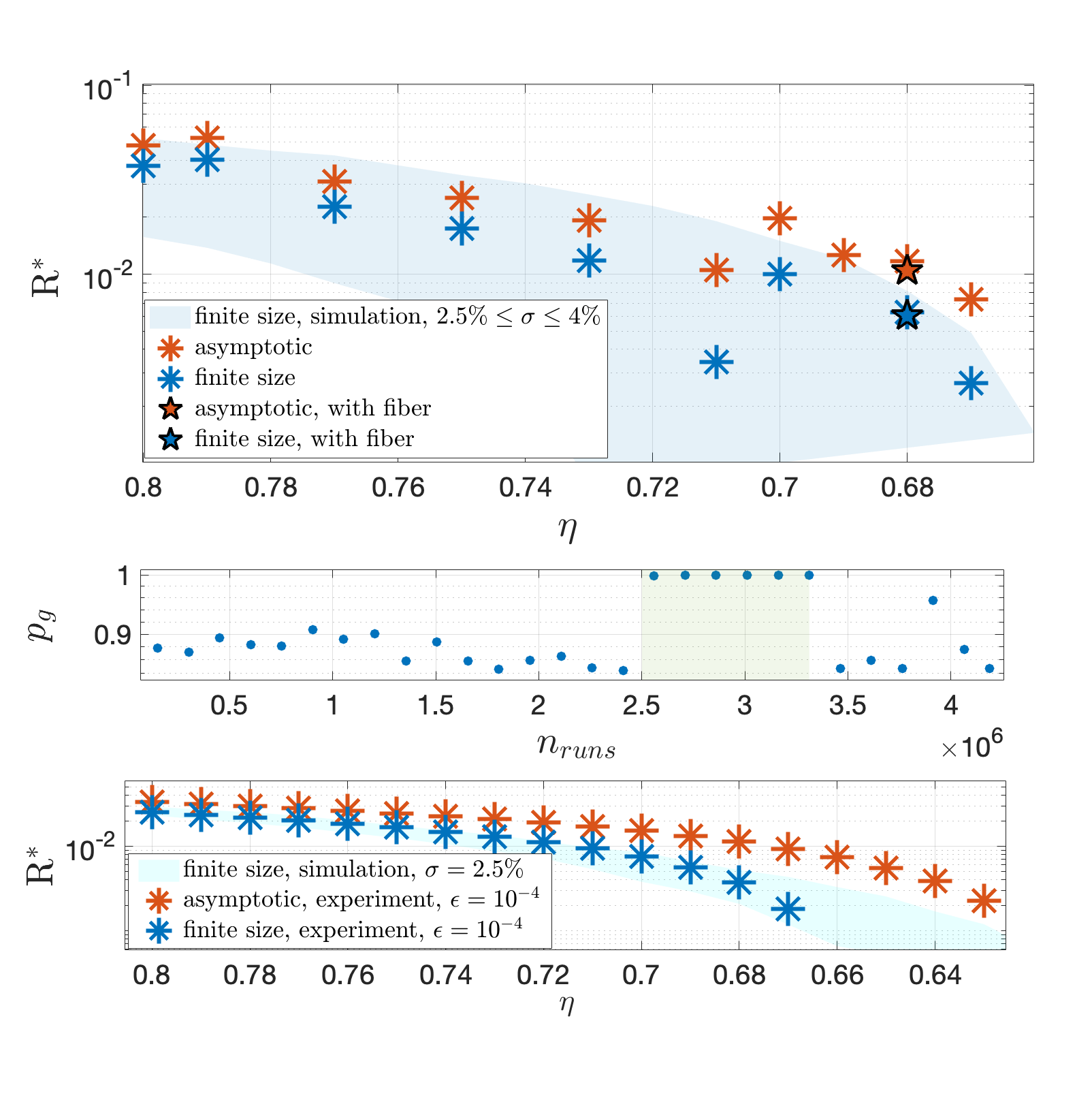}
	    \vspace{-13mm}
    		\caption{Experimental results. (Top) Key rate $R$ as a function of the transmission $\eta$ for the protocol with $n=2$ states. Each point represents a run over half an hour, with finite-size bound on the guessing probability (blue), see Appendix~\ref{app:finite} for details, and in the asymptotic regime (red). Data taken with $4.8\,\text{km}$ fiber corresponds to blue and red stars. Data is consistent with Monte Carlo simulations with polarization fluctuations $2.5\% \leq \sigma \leq 4 \%$ (blue region, estimated from data). (Middle) Illustration of the self-testing feature of the protocol. After 2.5 hours of operation, we artificially lower the detection efficiency of the SNSPD (shaded region), resulting in a guessing probability $p_g$ for Eve that jumps to one, hence $R=0$. Later, the SNSPD's efficiency is brought back to normal, hence $p_g<1$ again and $R>0$. (Bottom) Key rate $R$ vs transmission $\eta$ for the protocol with $n=3$ states, showing enhanced robustness to losses.}
    		\label{figResults}
	\end{figure}

The main challenge for the experimental implementation of our protocols lies in the limited loss budget ($<3\,\text{dB}$ for the 2-states protocol) of the quantum channel (QC), including Bob’s measurement setup.
For this feasibility experiment we used, for simplicity, weak coherent states, encoded and measured with fibered piezo-electric polarization controllers. Although these devices have low transmission loss, they encompass a limit in the operation speed ($\sim 1\,$kHz) (a much faster low-loss setup would be possible, though requiring a considerably higher degree of complexity). To minimize loss, on the detection side, we used a high efficiency SNSPD with 90\% detection efficiency at telecom wavelength and a dark count rate of $200\,\text{Hz}$ (ID Quantique). 

The experimental setup is shown on Fig.~\ref{fig:exp_setup}. On Alice’s side, a distributed feedback laser, triggered at a 1$\,$MHz rate, generates pulses at 1559$\,$nm with $90\,\text{ps}$ FWHM duration. The power fluctuation of the laser output signal is monitored every second using a powermeter. The rate of the optical signal is then reduced to 1$\,$kHz using an EOM. The polarization states $\ket{\phi_x}$ are encoded via a polarization controller (General Photonics' PolaRITE) comprised of 4 piezoelectric fiber squeezers. The polarization at the input of the controller is aligned so that two piezoelectric squeezers control the angles $\theta$ and $\phi$ respectively, allowing for the preparation of any desired state $\ket{\phi_x} = \cos{(\theta/2)}\ket{H} + \sin{(\theta/2)}\,e^{-i \phi_x}\ket{V}$ . In order to monitor the stability of the state preparation, part of the signal is sent to a monitoring setup that includes a manual polarization controller, a PBS, and a SPAD (ID Quantique ID210). 
 
Bob's measurement is implemented via a polarization controller (identical to Alice’s), allowing for an active choice of the measurement basis, followed by a PBS and the SNSPD.

The detection timestamps are recorded within a detection window of $1\,\text{ns}$, using a time-to-digital converter (ID Quantique ID800). 
All time driven components of the setup are triggered with an external clock (Silicon Labs SI5341). The inputs for Alice ($k$, $\mathbf{r}$) and Bob ($y$) are generated by a personal computer and transferred to the PPCs via Teensy micro controllers.

At the output of Alice, a VA controls the average photon number per pulse, $|\alpha|^2$, and when desired, introduces additional transmission loss. The QC, including Bob's setup and the detection efficiency, has a total transmission of $\eta=80.3\,\%\,(-0.953\,\text{dB})$. Adding $4815\,\text{m}$ of ultra-low loss telecom fiber (Corning SMF-28 ULL), the total transmission is $\eta=68\,\%\,(-1.674\,\text{dB})$. 

We first implemented the simplest protocol with $n=2$ states, investigating the performance as a function of the transmission of the QC, $\eta$. The two polarization states are prepared as in Eq.~\eqref{eq1} with $\theta = 0.6\,\text{rad}$. The coherent state $\alpha$ is optimized depending on $\eta$ (see Appendix~\ref{app:alpha}). The monitoring stage allows one to continuously measure the polarization and intensity of the light, in order to ensure the validity of the overlap assumption of Eq.~\eqref{eq3}.

To establish a secret key, we run each experiment for $0.5$ hours, collecting blocks of $N=1.8\times 10^6$ events. For each round, a random key bit $k=x$ is generated for Alice, and a random measurement setting $y$ for Bob. The detection (non-detection) at the SNSPD sets the outcome to $b=0$ ($b=1$). Once a block of data is collected, sifting is performed followed by a finite-size analysis to determine the statistics $p(b|x,y)$ (see Appendix~\ref{EvesInfoApp}). Given the overlap assumption and the statistics, we then compute a lower-bound on the key rate $R$ as in Eq.~\eqref{eq5} with the \gls{SDP} relaxation.

The results are shown in Fig.~\ref{figResults}~(Top); we also provide a bound on $R$ in the asymptotic regime, i.e.\ omitting finite-size corrections. To model the experiment, we perform a finite-size Monte Carlo simulation by ranging the estimated misalignment error $\sigma $ (see Appendix \ref{app:noise}) between $2.5\%$ and $4\%$ (estimated from data). Key rates in the order of $10^{-3}$ were obtained for a transmission of $\eta=67\%$. A similar rate is obtained over the $4.8\;$km fiber.

In Fig.~\ref{figResults} (Middle), we illustrate the  self-testing feature of the protocol. After $2.5\times10^6$ runs, the detector efficiency was intentionally reduced from $90\,\%$ to $42\,\%$ by lowering the bias current of the SNSPD. This sudden change is immediately detected in the post-processing, resulting in an immediate  increase of Eve's guessing probability to one. Hence, no secret key can be distilled and the users become aware of the setup's malfunction.

Lastly, we implemented the protocol with $n=3$ states (see Appendix \ref{3statesApp} for details). To lower misalignment errors (down to $\sigma=2.5\%$), due mostly to patterning effects in the polarization controllers, we implemented this protocol omitting the random choice of settings. Results, shown in Fig.~\ref{figResults} (Bottom), demonstrate increased robustness to loss, with positive key rates at $\eta=63\%$ in the asymptotic regime.

\section{Conclusion}

We presented protocols for receiver-DI QKD in a prepare-and-measure setup. Such a scenario is relevant in practice when Alice and Bob have a different level of trust in their devices; for instance Alice being a large company and Bob an end-user. Note that the MDI approach cannot provide security in the receiver-DI scenario, as some level of trust will always be required for both honest parties. We note that a detailed comparison of our approach with previous works is given in the companion paper \cite{Marie2}, as well as a more detailed discussion of how to justify the overlap bounds in practice.

From the observed statistics, the users can establish a secret key, while monitoring in real-time the correct operation of their devices, and thus immediately detect potential failure due, for instance, to a malfunctioning device or an attack by the eavesdropper (e.g., blinding). Our protocols do not rely on any type of fair-sampling assumption, contrary to Ref.~\cite{Tomamichel}.

The main limitation of our protocols are their loss sensitivity, which is in fact a fundamental limit for any QKD protocol in the receiver-DI scenario. Still, the requirements for transmission and efficiency are considerably relaxed compared to the full DI model (or the one-sided DI approach of Ref.~\cite{Branciard}), with the additional advantage that no source of entanglement is required, as well as significantly reducing the required efficiency on Bob's side (in principle, the efficiency can be made arbitrarily low, as we show in the companion article \cite{Marie2}).

We reported here a proof-of-principle implementation achieving security over a distance of few kilometers. We expect notable improvements in terms of rates and transmission for a high-speed polarization encoding setup as demonstrated in previous works \cite{grunenfelder2020}. Indeed, by multiplying the clock-rate by a factor of $10^3$, we would be much less limited by the block size and could increase the number of states of the protocol. For instance (following the results presented in Fig.~\ref{fig:exp_setup}) an implementation of the four-state protocol with weak coherent states encoded at a $1\,\text{GHz}$ rate, would lead to key rates of $\sim 100\,\text{kbps}$ at a $30\,\text{km}$ distance. Alternatively an implementation based on polarization-encoded single photons would also be more robust to loss \cite{Marie2}. 

To conclude, we believe that our work opens a new interesting approach in the intermediate regime between standard ``device-dependent'' QKD and the fully device-independent (DI) model, which is amenable to experiments. 

\emph{Acknowledgements.---}We thank Jonathan Brask and Denis Rosset for discussions, and ID Quantique for supplying the SNSPD. We acknowledge financial support from the EU Quantum Flagship project QRANGE, and the Swiss National Science Foundation (BRIDGE, project 2000021\_192244/1). This work was supported as a part of NCCR QSIT, a National Centre of Competence (or Excellence) in Research, funded by the Swiss National Science Foundation (grant number $51NF40-185902$).

\bibliographystyle{quantum}
\bibliography{myref}

\clearpage

% ...
\newpage

\appendix

\section{Bounding Eve's information} 
\label{EvesInfoApp}

In this appendix, we detail the procedure for obtaining upper bounds on Eve's guessing probability, which in turn leads to a lower bound on the key rate $R$.

The main assumption of the protocol is that the overlaps of Alice's prepared states $\ket{\psi_x}$ are limited. 
This is compactly expressed in terms of the Gram matrix
\begin{equation}\label{eq:Gram}
    G = \sum_{i,j=0}^{n-1} \gamma_{ij}\ketbra{i}{j}.
\end{equation}

Moreover, the observed statistics $p(b|x,y)$ are estimated by Alice and Bob. 
A lower bound on the asymptotic keyrate $R$ is given by
\begin{equation}
R = (H_\text{min}(p_g(e=k|\text{succ})) - H_2[\text{QBER}])p(\text{succ})
\end{equation}
where $p_g(e=k|\text{succ})$ is the probability that Eve guesses correctly the secret bit $k$ given that a round is conclusive, $H_\text{min}$ is the min-entropy, $H_2$ is the binary entropy function, QBER is the quantum bit error rate and $p(\text{succ})$ is the average probability to generate key. The QBER and $p(\text{succ})$ can be extracted from the observed statistics $p(b|x,y)$ while the guessing probability $p_g(e=k|\text{succ})$ needs to be upper bounded. The guessing probability is given by 
\begin{widetext}
\begin{equation}
\begin{aligned}
p_g(e=k|\text{succ}) &= \frac{p(e=k,\text{succ})}{p(\text{succ})} \\
&=\frac{\sum_{r=0}^{{n \choose 2}-1} p_R(r) \sum_{k=0}^1 p_K(k) \sum_{y=0}^{n-1} p_Y(y) \trace(\rho_{r_k}^{BE}M_{0|y} E_{k|r})(\delta_{y,r_0}+\delta_{y,r_1})}{\sum_{r=0}^{{n \choose 2}-1} p_R(r)\sum_{k=0}^1 p_K(k) \sum_{y=0}^{n-1} p_Y(y) \trace{(\rho_{r_k}^{BE}M_{0|y}\mathbb{1}) (\delta_{y,r_0}+\delta_{y,r_1}})},
\label{p_g_long}
\end{aligned}
\end{equation}
\end{widetext}
where $M_{b|y}$ are Bob's measurement operators with $b=0,1$ and $y=0,...,n-1$ and $E_{k|z}$ are Eve's measurement operators with $k=0,1$ and $z=0,...,{n\choose 2}-1$. $p_R(r)$, $p_Y(y)$ and $p_K(k)$ are the probabilities of choosing the inputs $r$, $y$ and $k$, satisfying $\sum_r p_R(r)=\sum_kp_K(k)=\sum_yp_Y(y) = 1$, $p_K(k)\geq0$ $\forall k$, $p_Y(y)\geq0$ $\forall y$ and $p_R(r)\geq0$ $\forall r$. Here, we take the input probabilities to be uniformly random for all inputs. As we impose no limit on the Hilbert space dimension, we can, without loss of generality (using Naimark's dilation theorem), assert that Bob's and Eve's measurements are projectors satisfying the following properties: $(i)$ $M_{b|y}M_{b'|y} = \delta_{b,b'}M_{b|y} \;\; \forall y,b,b' $, $(ii)$ $\sum_b M_{b|y} = \mathbb{1} \;\; \forall y$, $(iii)$ $E_{e|z}E_{e'|z} = \delta_{e,e'}E_{e|z} \;\; \forall z,e,e'$, $(iv)$ $\sum_e E_{e|z} = \mathbb{1} \;\; \forall z$ and $(v)$ $[M_{b|y},E_{e|z}]=0$ $\forall b,e,y,z $. The last property comes from the fact that Bob and Eve act on two different Hilbert spaces.

To upper bound $p_g(e=k|\text{succ})$, we will use the numerical method presented in \cite{Wang2019}. The numerical method consists of an \gls{SDP} hierarchy providing increasingly tight outer approximations of the set of quantum correlations in a discrete \gls{PM} scenario given the Gram matrix of the set of quantum states. 
This hierarchy provides a computationally tractable method to bounding $p_g(e=k|\text{succ})$ in the absence of any upper bound on the Hilbert space dimension.
Let us define the moment matrix $\Gamma$ of size $n q\times n q$:
\begin{equation}
\Gamma = \sum_{i,j=1}^n \Gamma_{ij} \otimes \ketbra{\hat{e}_i}{\hat{e}_j}
\end{equation}
where $\{\ket{\hat{e}_i}\}_{i=1,...,n}$ is an orthonormal basis in $\mathbb{R}^n$ and recall that $n$ is the number of states prepared by Alice. The sub-blocks $\Gamma_{ij}$ are defined as 
\begin{equation}
\Gamma_{ij} = \sum_{k,l=1}^r \bra{\psi_i}S_k^\dagger S_l \ket{\psi_j} \otimes \ketbra*{\hat{f}_k}{\hat{f}_l}
\end{equation}\\
 where $\{\ket*{\hat{f}_i}\}_{i=1,...,q}$ is an orthonormal basis for $\mathbb{R}^q$ and $\{S_i\}_{i=1,...,q}$ is a set of products of measurement operators $B_{b|y}$  and $E_{e|z}$. We note that the moment matrix $\Gamma$ is positive semi-definite and all the correlations $p(b,e|x,y,z)$ appear as elements in $\Gamma$.
One can choose the set of operators arbitrarily but the aim is to have as many independent operators as possible in the moment matrix. We can organize the operators into levels of the hierarchy. The first two levels are given by the two following sets of operators
 \begin{equation}
 \begin{aligned}
 \mathbb{S_1} &= \{\mathbb{1},B_{b|y},E_{e|z}\},\\
 \mathbb{S_2} &=  \mathbb{S_1} \cup \{B_{b|y}B_{b'|y'},E_{e|z}E_{e'|z'},B_{b|y}E_{e|z}\}.\\
 \end{aligned}
 \end{equation} 
One can go to higher levels by including increasingly long products of measurement operators. In Ref.~\cite{Wang2019}, it has been demonstrated that by going to increasingly large levels the hierarchy converges to the quantum set. 

The \gls{SDP} maximizing the guessing probability is given by 
\begin{equation}
\begin{aligned}
\max_{\Gamma} \;\;\;&p_g(e=k|\text{succ}) =  \frac{\trace(\Gamma A)}{\trace(\Gamma B)} &\\
\text{s.t.} \;\;\;&\trace(\Gamma C_{bxy}) = p(b|x,y)  &\forall b,y,x\\
&\trace(\Gamma D_{jk}) = G_{jk}  &\forall  j,k\\
&\trace(\Gamma F_{k}) = f_k\\
&\Gamma \geq 0
\end{aligned}
\end{equation}
where $A$ and $B$ are constant matrices selecting the term of the moment matrix necessary to compute the guessing probability Eq.~\eqref{p_g}. The fixed matrices $C_{bxy}$ set the appropriate entries of $\Gamma$ equal to the observed statistics $p(b|x,y)$, and $D_{jk}$ apply the inner-product constraint of the encoding states to $\Gamma$ (recall that $G$ is the Gram matrix~\eqref{eq:Gram} specifying these overlaps). Finally, $F_{k}$'s set the linear constraints arising from the operators $B_{b|y}$ and $E_{e|z}$. As written above, the optimization has the form of a semidefinite-fractional program, i.e., a semidefinite program with an objective function consisting of the fraction between two linear functions. Such a program can in general be transformed into a regular semidefinite program by applying a Charnes-Cooper transform as described in the theory of linear-fractional programming \cite{Charnes1962}.

However, in our case we notice that the denominator $\trace(\Gamma B)=p(\text{succ})$ only involves directly observable quantities, i.e., terms of the $\Gamma$ matrix which are fixed by the first constraint. Therefore, it is sufficient to simply optimize 
\begin{equation}\label{eq:SDP1}
\begin{aligned}
\max_{\Gamma} \;\;\;&p_g(e=k,\text{succ})=\trace(\Gamma A) &\\
\text{s.t.} \;\;\;&\trace(\Gamma C_{bxy}) = p(b|x,y)  &\forall b,y,x\\
&\trace(\Gamma D_{jk}) = G_{jk}  &\forall  j,k\\
&\trace(\Gamma F_{k}) = f_k\\
&\Gamma \geq 0.
\end{aligned}
\end{equation}

The \gls{SDP} \eqref{eq:SDP1} assumes the overlap assumption is satisfied exactly, but this can easily be relaxed as follows.
For $n=2$ we can, without loss of generality, consider real overlaps and hence enforce simply a lower bound on the overlaps, i.e., $\gamma=\braket{\psi_0}{\psi_1}\geq C$. For $n>2$, we assume that the the real and imaginary part of the overlap are in the vicinity of the ideal values $\gamma_{ij}$ by upper and lower bounding the real and imaginary parts of $\gamma_{ij}$
\begin{equation}
    \begin{aligned}
      |\text{Re}(\gamma_{ij}) - \text{Re}(\braket{\psi_i}{\psi_j})| &\leq \epsilon \\
      |\text{Im}(\gamma_{ij}) - \text{Im}(\braket{\psi_i}{\psi_j})| &\leq \epsilon.
    \end{aligned}
\end{equation}

\subsection{Finite-size statistics}
\label{app:finite}
The primal \gls{SDP} (\ref{eq:SDP1}) gives optimal bounds on the joint guessing probability given a Gram matrix and an asymptotic probability distribution, but it is not practical when we consider finite-size statistical effects. Indeed, finite experimental data does not describe, in general, asymptotically valid distributions, and it is not clear \emph{a priori} how finite-size effects propagate in the primal \gls{SDP}. A general way to deal with the first issue in NPA-type SDP hierarchies was proposed in~\cite{Lin18}. In our case, since we are simply interested in obtaining a valid bound on the keyrate which applies to the finite statistics, we rely only on the symmetrized observed statistics $p(b|x,y)$. We then analyse the effect of finite statistical fluctuations with the help of the dual of the \gls{SDP}. Namely, we will first show that the dual objective function upper bounds the primal objective function, and then we will upper-bound the objective function of the dual by taking into account the finite-size statistical effects. Finally, we will lower-bound $p(\text{succ})$.

The primal \gls{SDP} has the generic form
\begin{subequations}
\begin{align}
\max_{\Gamma} &\trace(\Gamma A)&\label{objfunc1}\\
\text{s.t.} \;\;\;
&\trace(\Gamma B) = b \label{subeqprimal1}\\
&\Gamma \geq 0. \label{subeqprimal2}
\end{align}
\end{subequations} 
To derive the dual, we introduce for each constraint a Lagrangian multiplier $\beta$, and $H$. The Lagrangian function of the primal \gls{SDP} is given by
\begin{equation}\label{lagrangian}
\mathcal{L} = \trace(\Gamma A) + \beta(b-\trace(\Gamma B)) + H \Gamma.
\end{equation}
We define $\mathcal{S}$ to be the supremum of the Lagrangian over the primal \gls{SDP} variable in its domain:
\begin{equation}
\mathcal{S} = \sup_{\Gamma} \mathcal{L}.
\end{equation}
Examining this quantity, we notice that the contributions coming from the second term of the Lagrangian (\ref{lagrangian}) vanish because of the constraint \eqref{subeqprimal1}. Imposing $H\geq0$ implies that the second term of the Lagrangian is positive. Since the first term of the Lagrangian is equal to the objective function of the primal \gls{SDP} (\ref{objfunc1}), $\mathcal{S}(\beta,H)$ is an upper bound on the objective function of the primal \eqref{objfunc1} whenever $H\geq0$, i.e.\ $\mathcal{S}\geq p_g^\text{primal}$. Let us rewrite $\mathcal{S}$ by grouping the primal \gls{SDP} variables 
\begin{equation}
\mathcal{S} = \sup_{\Gamma} \trace(\Gamma(A-\beta B+H)) + \beta b.
\end{equation}
To obtain an optimal upper bound on $p_g^\text{primal}$ from $\mathcal{S}$, we minimize $\mathcal{S}$ over the Lagrange multipliers. Since $\Gamma\geq0$ and $H\geq 0$ the supremum may be unbounded if the first term doesn't vanish. Hence, we impose $A-\beta B+H=0$. This leads to the following \gls{SDP}
\begin{subequations}\label{eq:dual}
\begin{align}
p_g^\text{dual}=\min_{\beta} \;\;\;  &\beta b\\
\text{s.t.} \;\;\;
& A-\beta B \leq 0.
\end{align}
\end{subequations}
By construction, we thus showed that  $p_g \leq p_g^\text{primal} \leq p_g^\text{dual}$.

With the above considerations it is easy to show that the objective function of the dual is given by
\begin{equation}
p_g \leq K + \sum_{x,y} \nu^{b=0}_{x,y} \, p(b=0|y,x),
\label{dualObjFunction}
\end{equation}
where $K$ takes into account all the terms that do not contain any finite-statistical effects and $\nu^{b=1}_{x,y}$ is the associated set of dual variables corresponding to the constraints imposed by the probability distribution. Let us upper bound the second term of Eq.~\eqref{dualObjFunction}. We first write
\begin{align}
\sum_{x,y} \nu^{b=0}_{x,y} p(b=0|x,y) &=\sum_{x,y} \frac{ \nu^{b=0}_{x,y}}{p(x)p(y)}\delta_{b,0}^{x,y}p(b,x,y) \notag\
&= \sum_{x,y} g^{b=0}_{x,y} \mathbb{E}(\delta_{b,0}^{x,y}).
\label{eqbound}
\end{align}
Here, $\delta_{b,0}^{x,y}$ can be interpreted as a binary game which is won if $b=0$ and lost if $b=1$. If the game is won, we score $g^{b=0}_{x,y}:= \frac{\nu^{b=0}_{x,y}}{p(x)p(y)}$. Then, using Theorem B.2 of Ref.~\cite{Bancal2018}, we obtain the following bound on $\mathbb{E}(\delta_{b,0}^{x,y})$ with confidence $1-\alpha_1$: $P(\mathbb{E}(\delta_{b,0}^{x,y}) \leq \hat q_{x,y}) \geq 1-\alpha_1$ for $\hat q_{x,y} = 1-I_{\alpha_1}^{-1}(f(b=0,x,y) N,N(1-f(b=0,x,y))+1)$ where $I_{\alpha}^{-1}(a,b)$ denotes the inverse regularized Beta function and $\alpha_1 = 10^{-9}$. We thus obtain the following bound on Eq.~\eqref{eqbound} with confidence $1-\alpha_1$:
\begin{equation}
\sum_{x,y} g^{b=0}_{x,y} \mathbb{E}(\delta_{b,0}^{x,y}) \leq \sum_{x,y} g^{b=0}_{x,y} \hat q_{x,y}.  
\end{equation}
Using Eq.~\eqref{dualObjFunction} this translates into a bound on $p_g$ with the same confidence level $1-\alpha_1$:
\begin{equation}
p_g \leq  K + \sum_{x,y} g^{b=0}_{x,y} \hat q_{x,y}.   
\end{equation}
We hence upper-bounded the joint guessing probability \eqref{eq:SDP1}. It remains to lower-bound $p(\text{succ})\geq p_2^\ast$.

The probability of success is given by
\begin{equation}
\begin{aligned}
p(\text{succ}) &= \sum_{r,k,y} \frac{(\delta_{y,r_0}+\delta_{y,r_1})}{p_R(r)p_K(k)p_Y(y)} p(b=0,r_k,y)\\
&=\sum_{r,k,y} g_{r_k,y}^{b=0}(\delta_{y,r_0}+\delta_{y,r_1})\delta_{b,0}p(b,r_k,y)\\
&=\sum_{r,k,y} g_{r_k,y}^{b=0}\mathbb{E}(\chi(b,r_k,y))\\
&\geq \sum_{r,k,y} g_{r_k,y}^{b=0}\hat q_{r_k,y}.
\end{aligned}
\end{equation}
Similarly as before, $\chi(b,r_k,y):=(\delta_{y,r_0}+\delta_{y,r_1})\delta_{b,0}$ can  be interpreted as a binary game which is won if $b= 0$ and $y=r_1$ or $y=r_0$ and lost if $b=1$. If the game is won, we now score $g_{r_k,y}^{b=0}=\frac{1}{p_R(r)p_K(k)p_Y(y)}$. We obtain the following lower bound on $\mathbb{E}(\chi(b,y,r_k))$ with confidence $1-\alpha_2$: $P(\mathbb{E}(\chi(b,r_k,y) \geq \hat q_{r_k,y}) \geq 1-\alpha_2$ for $\hat q_{r_k,y} = I_{\alpha_1}^{-1}(f(b=0,r_k,y) N,N(1-f(b=0,r_k,y))+1)$.

Since $p(e=k,\text{succ})\leq p_1^*$ with prob $1-\alpha_1$ and $p(\text{succ})\geq p_2^*$ with prob $1-\alpha_2$, we deduce that $p(e=k|\text{succ})\leq \frac{p_1^*}{p_2^*}$ with confidence at least $1-\alpha$, for $\alpha=\alpha_1+\alpha_2$.

With the upper bound on the guessing probability $p(e=k|\text{succ})\leq \frac{p_1^*}{p_2^*}$ taking into account finite statistical effects, we estimate the key rate in Eq.~\eqref{eq5} as
\begin{equation}
R^*\geq \left(H_\text{min}\left(\frac{p_1^*}{p_2^*}\right) - H_2[\text{QBER}]\right)p(\text{succ}).
\end{equation}

\section{Noise model}
\label{app:noise}
In this section we are going to consider a channel model that allows us to estimate the amount of noise in our experiment due to misalignment between the states prepared by Alice and the bases chosen by Bob. These errors can be divided in two different categories: systematic errors and stochastic ones. 

In order to model the former it is sufficient to add a constant term of misalignment for both the angle $\theta$ and the state and basis choice $x$ and $y$. If we define these terms as $\Delta_{\theta}$ and $\Delta_{x,y}$ we can rewrite Eq.~\eqref{eq4} as:
\begin{equation}
p(b=0|x,y) = 1-e^{-|\alpha|^2\sin(\theta + \Delta_{\theta})^2\sin(\frac{\pi(x-y+\Delta_{x,y})}{2})^2}.
\end{equation}

If we now consider instead possible stochastic errors we have to assume a certain noise distribution of the polarization inside the fiber. In this scenario, we analyze the case of polarization fluctuation, which average state is aligned with the prepared one. In order to do so we can divide our channel in different steps. First, we consider our source to generate a coherent state with mean photon number $\mu = |\alpha|^2$ in the polarization $H$:
\begin{equation}
         \ket{\psi}  = \ket{\alpha}_H  \ket{0}_V.
\end{equation}
Alice then turns this state into the prepared one shown in Eq.~\eqref{eq1} by a unitary transformation $A_{\theta,x}$ that turns the polarization into the desired one. The state propagates into the channel that can be divided into a transformation that represents the loss, i.e., $C_{\eta}$ and one that represents the polarization fluctuations that we want to model in this section, i.e., $C_{\theta,\phi}$ where $\theta$ and $\phi$ are the random variables representing respectively the polar and azimuthal rotation on the Poincaré sphere. Bob chooses the measurement basis by an analogous transformation of Alice, i.e.\ $B_{\theta,y}$. Finally, the two polarization modes are separated by a polarizing beam splitter in two paths (as shown in Fig.~\ref{fig:exp_setup}). One of the two paths corresponds to the projection needed and leads to a single photon detector. The resulting state before the PBS has the form:
\begin{equation}
    A_{\theta,x}C_{\theta,\phi}C_{\eta}B_{\theta,y}\ket{\alpha}_H  \ket{0}_V = \ket{\beta_{0|\theta,x,y,\theta,\phi}}_{H'}  \ket{\beta_{1|\theta,x,y,\theta,\phi}}_{V'},
\end{equation}
where $H'$ and $V'$ are the proper polarization modes of Bob's PBS and $\beta_{1|\theta,x,y,\theta,\phi}$ is the coherent state amplitude arriving on the single photon detector whilst $\beta_{2|\theta,x,y,\theta,\phi}$ is the amplitude that correspond to the other output port of the PBS.

The transformation we are interested in is $C_{\theta,\phi}$, since this corresponds to our source of misalignment. Without loss of generality, since the rotation acts randomly and independently on each state, we can consider the effect of this transformation already in the input state of the system giving us a new input state
\begin{equation}
    \ket{\psi'}  = \ket{\alpha \cos(\theta/2)}_H  \ket{\alpha \sin(\theta/2)e^{i\phi}}_V.
\end{equation}

For simplicity we consider $\theta$ to be a random variable with a normal Gaussian distribution, i.e.\ $P(\theta) = \frac{1}{\sigma_{\theta}\sqrt{2\pi}} e^{-\frac{\theta^2}{2}\sigma_{\theta}^2}$ with null mean and standard deviation $\sigma_{\theta}$, and $\phi$ to be uniformly distributed in the interval $[0,2\pi]$. In order to show the effect of this fluctuation on the detection statistics, it is sufficient to focus on the probability of having a conclusive event (corresponding to $b = 0$). This probability has the form $p(b=0|x,y) = 1-e^{-|\beta_{0|\theta,x,y}|^2}$, where $|\beta_{0|\theta,x,y}|^2$ is the mean photon number of the coherent state arriving to the detector in the configuration where $\theta$ is fixed and Alice and Bob chose their inputs as $x$ and $y$ respectively.

By trivial calculation using $\ket{\psi'}$ as initial state, we can express the value of $|\beta_0|^2$ depending on the specific values of $\theta$ and $\phi$ as follows:
\begin{equation}
\begin{aligned}
    &|\beta_{0|\theta,x,y,\theta,\phi}|^2 = \\
    &[\cos^2(\theta/2)(r^2+t^2+2rt\cos(x-y)) \\
    &+ \sin^2(\theta/2)2rt(1-\cos(x-y))\\
    &- \cos(\theta/2)\sin(\theta/2)[(t-r)2\cos(\phi)\\
    &-2\cos(\phi+x-y)+4r\cos(\phi)\cos(x-y)]]|\alpha^2|,
\end{aligned}
\end{equation}
where $r = \sin(\theta/2)^2$ and $t = \cos(\theta/2)^2$.

Finally, it is sufficient to take the average of this quantity with respect to the probability density function of $\theta$ and $\Phi$ in order to obtain the final mean photon number $|\beta_{0|\theta,x,y}|^2$:
\begin{equation}
\begin{aligned}
    |\beta_{0|\theta,x,y}|^2 =& \int_0^{2pi}\int_{-\infty}^{+\infty}P(\theta)U(\phi)|\beta_{0|\theta,x,y,\theta,\phi}|^2d\phi d\theta \\
    =& [\frac{1}{2}(1+e^{-2\sigma_{\theta}^2})(r^2+t^2+2rt\cos(x-y)) \\
    & +\frac{1}{2}(1-e^{-2\sigma_{\theta}^2})2rt(1-\cos(x-y))]|\alpha|^2.
\end{aligned}
\end{equation}

In conclusion, this simple model allows us to evaluate the stability and the fluctuation of our channel just by a simple figure of merit given by the standard deviation $\sigma_{\theta}$. It is important to stress that this calculation shows just a possible channel model between Alice and Bob for simulation purposes and are not needed to grant the security of the protocol.

\section{Experimental implementation}
\subsection{Mean photon number}
\label{app:alpha}
\begin{figure}[h!]
		\centering
        \includegraphics[width=0.8\linewidth]{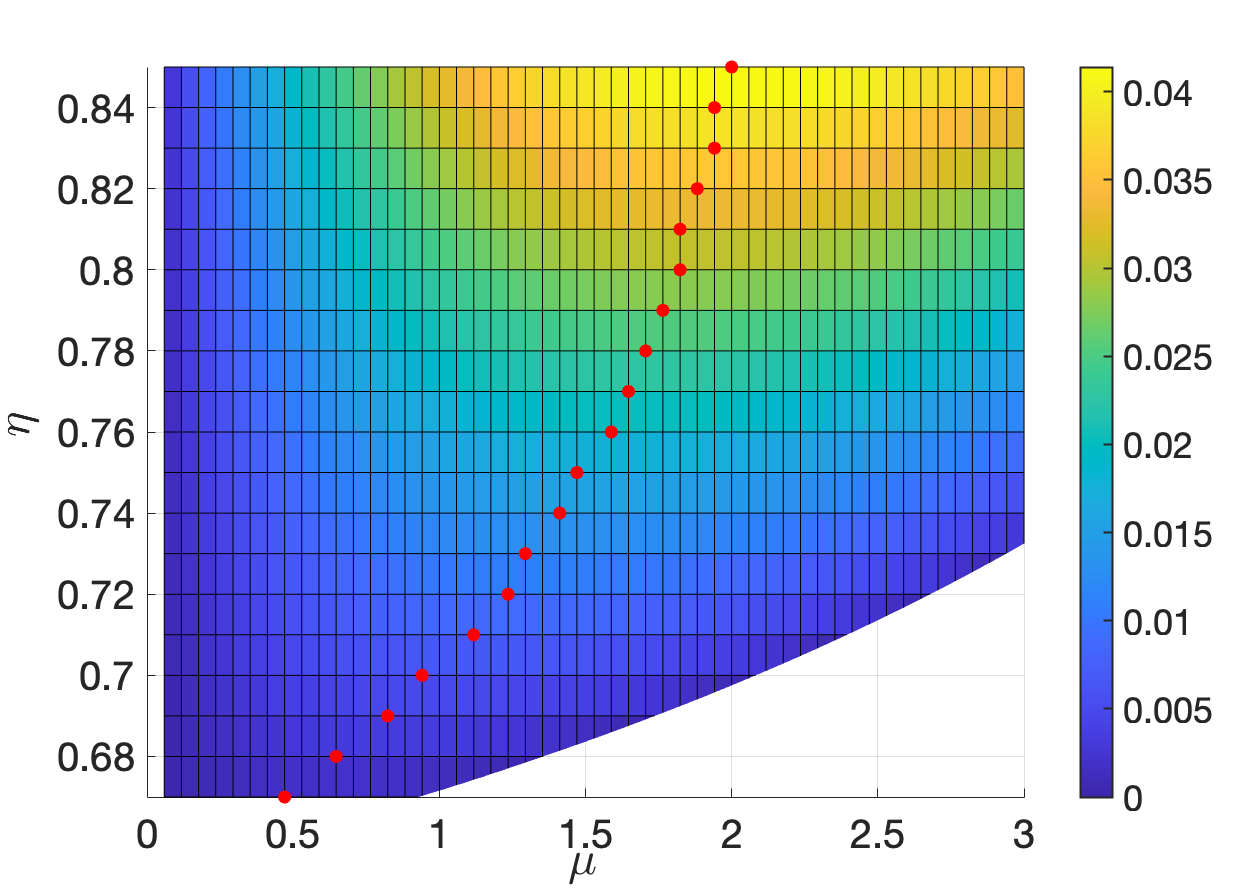}
    	\caption[Results.]{$n=2$. Key rate $R$ as a function of $\eta$ and $\mu=|\alpha|^2$ for $\theta=0.6$ rad, $\sigma=4\%$ and $p_{dc}=3.24\cdot10^{-7}$. The red dots indicate the optimal value of $|\alpha|^2$ for each $\eta$.}
    	\label{fig:optmu}
\end{figure}

Here we discuss the question of how to choose the polarizations and coherent state amplitude $\alpha$ for the implementation. For a given transmission $\eta$, we optimized numerically the key rate $R$ over $\alpha$ and $\theta$ (note that the second angle $\phi$ is already defined, and depends on the number of states $n$). To do so, we also take into account the probability of having a dark count (estimated to $p_{dc}=3.24\cdot10^{-7}$) and the misalignment error (pessimistically estimated to $\sigma=4\%$; see Appendix~\ref{app:noise}). For practical convenience, we then decided to fix $\theta$ in order to be able to swipe automatically over $\alpha$ while running the experiment and not having to realign manually the polarization states for each transmission. We choose $\theta$ so that we can get key for a large span of transmissions $\eta$. For every $\eta$ and fixed $\theta=0.6$ rad, we extract the optimal $\alpha$. In Fig.~\ref{fig:optmu}, we show the numerical optimization for $n=2$.

\subsection{Illustration of the self-testing feature}
Fig.~\ref{figResults} (Middle) shows the evolution of the guessing probability $p_g(e=k|\text{succ})$ throughout the measurement's duration for the two-state protocol implementation with the $4.8\,\text{km}$ optical fiber. After $2.5\times10^6$ runs, the detector efficiency was intentionally reduced from $90\,\%$ to $42\,\%$ by lowering the bias current of the SNSPD. This was done to mimic a malfunction or tampering by Eve of the detector. The bias current was kept low until $3.5\times10^6$ runs had elapsed. During this time, the guessing probability rises up to $p_g=1$, hence no key can be extracted. The parties become aware of a deviation from the correct operation regime, and can then abort the protocol and re-calibrate their devices.

\subsection{Implementation of the protocol with $n=3$ states}
\label{3statesApp}
For the 3-state protocol we set the polarizations to $\theta = 0.7\,\text{rad}$; and $\phi=\frac{2}{3}\pi$ as defined in the presentation of the protocol. 
Because of the limiting patterning effect of the polarization controller in this scenario, the measurement was performed without the random basis choice, i.e., using fixed states and measurement basis. The detection rate was kept at $1\,\text{kHz}$ but the states were switched after $50\,\text{ms}$, following a fixed sequence. As a result, we observe a reduction of the the misalignment error down to $\sigma=2.5\%$. The measurement was run with an intensity $|\alpha|^2=0.647$ and transmission $\eta=65\%$ (setting the variable attenuator at the output of Alice's device), integrating over a period of 2 hours. As the variable attenuator sets both the intensity of the prepared states and the transmission of the channel, given by the product $|\alpha|^2 \eta = c$, the data can in fact be analyzed in different ways.  Specifically, we can consider a value of $\eta$ between $80\%$ and $60\%$. The corresponding value of the intensity is then given by $|\alpha|^2 = c/\eta$. We observed a positive secret key rate for $\eta$ as low as $63\%$ and $67\%$ for the asymptotic and finite-size analysis, respectively (see Fig.~\ref{figResults}~(Bottom)). To check the consistency of the data, we performed a Monte Carlo simulation, with finite-size analysis (assuming errors $\sigma=2.5\%$). The expected range is given by the blue area, and all data points are inside.

\end{document}